\documentclass[pre,twocolumn,amsmath,showpacs,floatfix,superscriptaddress]{revtex4}
 \usepackage{graphicx}

\begin{document}
\title{Epidemic processes with immunization}

\author{Andrea Jim\'enez-Dalmaroni}
\email{jimenez@thphys.ox.ac.uk}
\affiliation{University of Oxford, Department of Physics --
        Theoretical Physics, 1 Keble Road, Oxford OX1 3NP, U.K.}
\author{Haye Hinrichsen}
\affiliation{Theoretische Physik, Fachbereich 8, Bergische
Universit{\"a}t Wuppertal, D-42097 Wuppertal, Germany}

\begin{abstract}
We study a model of directed percolation (DP) with immunization, i.e.
with different probabilities for the first infection and subsequent infections. The immunization effect leads to
an additional non-Markovian term in the corresponding field theoretical action. We
consider immunization as a small perturbation around the DP fixed
point in $d<6$, where the non-Markovian term is relevant. The
immunization causes the system to be driven away from the neighbourhood of
the DP critical point. In order to investigate the dynamical critical
behaviour of the model, we consider the limits of low and high first
infection rate, while the second infection rate remains
constant at the DP critical value. Scaling arguments are applied to obtain an expression for the survival
probability in both limits. The corresponding exponents are written in
terms of the critical exponents for ordinary DP and DP with a wall. We find that the survival probability does not obey a
power law behaviour, decaying instead as a stretched exponential in
the low first infection probability limit and to a constant in the high
first infection probability limit. The theoretical predictions are confirmed by optimized numerical
simulations in $1+1$ dimensions.
\end{abstract}

\maketitle
\def\xvec{{\text{\bf x}}}
\parskip 2mm

\section{Introduction}
\label{Introduction}

Epidemic processes can be described as the spread and decay of a
non-conserved agent, an example of which is an infectious
disease~\cite{Mollison77}. The agent is not allowed to appear
sponstaneously but it can multiply itself by infecting neighbouring
individuals, or decay at a constant rate. Depending on the balance
between these two processes, the infection may either die out or spread over the entire
population. The two regimes of survival and extinction of
the epidemic are typically separated by a continuous non-equilibrium
phase transition. When the decay process dominates, the epidemic dies out at large times and the
system gets trapped in an \textit{absorbing state} from which it cannot escape.

Continuous phase transitions into absorbing states are associated
with certain universality classes~\cite{MarroDickman,Hinrichsen00}. For epidemic processes, a
well studied case is the universality class of directed percolation
(DP). It is believed that two-state spreading processes with
short-range interactions generically belong to the DP class, provided that
quenched randomness, unconventional symmetries and large scales due to
memory effects are absent~\cite{Janssen81,Grassberger82}. Examples of
physical systems whose critical behaviour is described by DP
include heterogeneous catalysis~\cite{ZGB86}, chemical reactions~\cite{GdelaT79,Schlogl72},
interface depinning~\cite{TangLeschhorn92, Buldyrev}, the onset of
spatio-temporal chaos~\cite{RRR03}, flowing sand~\cite{FlowingSand} and self-organized criticality~\cite{MohantyDhar}.

The epidemic process in which the susceptibility to infection is
independent of previous infections is described by DP. However, for a
more realistic description, we should consider an {\em immunization}
effect~\cite{Perelson97}. Immunization can be added to the DP model by
changing the susceptibility after the first infection
~\cite{Cardy83,CardyGrassberger85,Janssen85}. A minimalistic model
that captures this feature is one that is controlled by two independent
parameters: a probability of first infection and another
probability for all subsequent reinfections. The fact that the
local susceptibility depends on whether a site has been infected in
the past or not leads to a {\em non-Markovian} epidemic process, in
which the time evolution depends on the entire history. This
non-Markovian feature changes the universality class of the epidemic spreading.

\begin{figure}
\includegraphics[width=85mm]{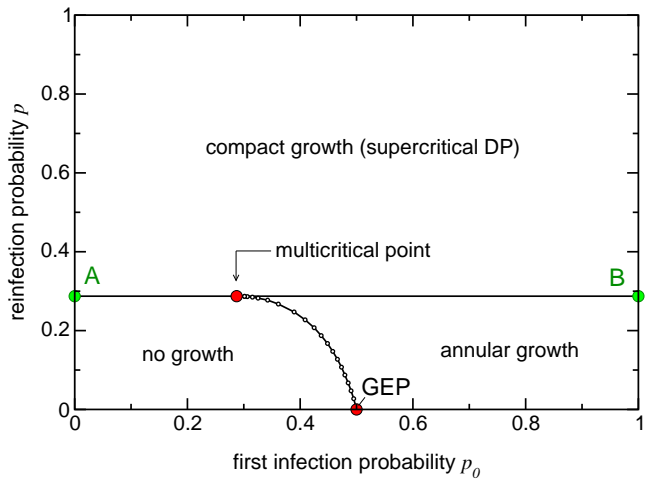}
\caption{
\label{FIGPHASEDIAG}
Phase diagram of directed bond percolation with immunization in 2+1
dimensions. Along the curve phase transition line the universality
class corresponds to the GEP. The horizontal line separates the no
growth-annular growth region from the compact growth behaviour. The
point where both phase transitions lines meet correspond to the universality class of DP. 
} 
\end{figure}

The phase diagram of an epidemic process with immunization
(Fig.~\ref{FIGPHASEDIAG}) was studied in Ref.~\cite{GCR97, MGD98}. If the
probabilities for first infections and reinfections are equal, the model corresponds to ordinary directed
percolation. However if the susceptibility changes to zero after the first
infection there is \textit{perfect} immunization, and the model reduces to the General Epidemic Process
(GEP)~\cite{Mollison77}. GEP belongs to the ordinary percolation
universality class~\cite{Alexandrowicz80}. The critical points of the GEP and DP
are connected by a curved phase transition line separating a phase in
which the spreading process always dies out, from another phase of
annular growth, where an active front may propagate into regions of
non-immune sites, leaving a bulk of immune sites behind. As shown in
Ref.~\cite{GCR97} the critical behavior along this line (except for
the upper terminal point) belongs to the same universality class as
GEP. Using field theoretic renormalization group techniques,
the critical exponents were calculated along this line ~\cite{CardyGrassberger85,Janssen85}.
The main result in Ref.~\cite{GCR97} is that the compact growth/no
growth phase transition line is at the critical value of the
reinfection rate and independent of the first infection rate.
Above this horizontal transition line in Fig.~\ref{FIGPHASEDIAG}, the model
exhibits compact growth and approaches the stationary
state of supercritical DP. This is because, in the active phase of an epidemic process with immunization, each site 
will be visited at least once after a sufficiently long time so that the
dynamics in the stationary active state involves only
reinfections. On the horizontal phase transition
line itself all reinfection processes are critical DP while the
probability of first infections may be sub- or supercritical. By
varying the first infection rate, we can impede or facilitate the spreading into
non-immune regions. In Ref.~\cite{GCR97} a numerical analysis of the
scaling behaviour along this horizontal line gives no hint of power
law behaviour, and it is also suggested that this result could be
applied to models with multiple absorbing
states~\cite{MGD98, JensenDickman93, MDHM94, Dickman96, Munoz96, LopezMunoz97, Munoz97, HwangPark99,FvW02}.

The aim of the present work is to investigate in further detail the dynamical critical behavior along
the horizontal phase transition line and in the vicinity of the DP
critical point, and to give a theoretical explanation of the absence of power law scaling 
along this line.

In order to describe the effect of immunization in an epidemic process, in Sec.~\ref{FT} we study a field theoretic
formulation of directed percolation with immunization and we show that the non-Markovian term
which contains the immunization effects, is relevant under renormalization group analysis. As a consequence of
this result we argue that the asymptotic spreading properties along
the horizontal transition line should be determined by the limits of
very low and very high first infection probability. Therefore, in
Sec.~\ref{LOW} we
present a study of the very small infection probability limit and
we develop a quasi-static approximation to obtain the scaling
behaviour of the survival probability of the epidemic. It
turns out that this does not follow a power law, but instead decays
asymptotically as a stretched exponential. This theoretical prediction
is confirmed by optimized numerical simulations in 1+1 dimensions. The high first infection probability limit is studied in
Sec.~\ref{HIGH} giving similar results.  We
complete the section with a theoretical
approximation for the spreading behaviour, and with numerical calculations
to corroborate the
theoretical claims. Finally, a discussion of
these results together with  a possible connection with multiple
absorbing state models is the subject of the conclusions in Sec.~\ref{Conclusions}. 
%
%
%
\section{Field theoretical analysis}
\label{FT}

\subsection{The model}
\label{Action}
%
\noindent
In this section we develop an alternative derivation of the action for
the DP model with immunization that was proposed
in~\cite{Cardy83,CardyGrassberger85,Janssen85}. The microscopic rules
for DP in $d+1$ dimensions are rather simple. An infected site at time
$t$ can infect its nearest neighbours at time $t+1$ with a
probability $p_0$. There is a critical threshold $p_c$ such that for $p_0<p_c$ the epidemic
process always dies out, that is, it reaches the absorbing state. For
$p_0>p_c$ there is a finite probability that the epidemics
survives. At the critical point $p_0=p_c$ the system scales
anisotropically in time and space. The  upper critical dimension is
$d_c=4$, below which the fluctuation effects become important. The 
field theoretic action of
DP~\cite{GrassbergerSundermeyer78,CardySugar80} reads as follows:
\begin{equation}
\label{DPAction}
S_{\rm DP} = \int dt \, d^{d}x \bigl[ \tilde{\phi}( \partial_t - D
\nabla^{2} + r) \phi + u_1\tilde{\phi}\phi^{2} -
u_{2}\tilde{\phi}^{2}\phi\bigr]\, .
\end{equation}
Here, $\phi$ is the local activity, $\tilde{\phi}$ is the response
field and $r \propto p_c-p_0$ is the mass parameter which measures the distance from criticality.
This action can also be written as a Langevin-type equation for the
local activity,
\begin{equation}
\label{DPLangevin}
( \partial_t - D \nabla^{2} + r) \phi + \frac{1}{2} u \phi^{2} +
\xi(\mathbf{x},t)=0 \,,
\end{equation}
 where $u$ is the symmetrized coupling constant after rescaling the
fields according to the DP time reversal symmetry. The noise
$\xi(\mathbf{x},t)$ is Gaussian, and satisfies $\langle \xi(\mathbf{x},t) \rangle =0 $ and
$\langle \xi(\mathbf{x},t) \xi(\mathbf{x}',t')\rangle = u \delta^d(\mathbf{x}-\mathbf{x}')
\delta(t-t')$. For a systematic analysis of the
immunization around the DP fixed point it is often more convenient to make use of the 
description in terms of the action.

To add immunization to the model, the susceptibility
after the first infection is decreased by an amount $\lambda>0$. The probability of
infection, $p$, depends locally on position and time, $p=p(\mathbf{x},t)$. The microscopic rules 
are modified as follows: a \textit{healthy} site can first be infected with
infection probability $p_0$. A site which is infected at time $t$ will
become immune at the next time step $t+1$. Meanwhile any immune site can be re-infected at the
lower re-infection probability $p_0-\lambda$. Thus, the state of the system
at time $t$ depends not just on the configuration of infected sites at time $t-1$, but
on the entire previous history. 

In order to take into account the effects of
immunization, we will modify the DP action. Let us consider a discrete $d+1$
dimensional lattice and assume that the field
$\phi(\mathbf{x},t)$ is almost constant between
times $t$ and $t+\Delta t$, where $\Delta t=1$ is the time step unit on the lattice. We
then subdivide it into $N$ intervals, $\Delta t/N$. First of all, we
want to find an expression 
for the probability that there is an active site
in the temporal sub-interval $\Delta t/N$. To do so in terms of the field  $\phi(\mathbf{x},t)$,
we need to take into account the fact that $\phi$ has units of
$length^{-d/2}$. Then this probability can be written
as $w \phi(\mathbf{x},t)\Delta t/N $, where we introduce a parameter $w$
to ensure that this expression is dimensionless.

In this way, $1- w \phi(\mathbf{x},t) \frac{\Delta t}{N}$ is the probability that the
site is not active in the interval $\Delta t/N$. The
probability of not finding activity between times $t$
and $t+ \Delta t$, can then be expressed as:

\begin{eqnarray}
\prod_{i=0}^{N} ( 1- w\phi(\mathbf{x},t+i/N) \frac{\Delta t}{N}) &
\sim & (1- w\phi(\mathbf{x},t) \frac{\Delta t}{N})^{N} \nonumber \\
& \stackrel{N \rightarrow \infty}{\longrightarrow} & e^{-w\phi(\mathbf{x},t) \Delta t}.
\end{eqnarray} 

However, between times $t=0$ and $t>1$, we can no longer assume
that the field $\phi$ is constant. Therefore, the
probability of a site $\mathbf{x}$ has never been infected by time $t$ turns out to be,

\begin{equation}
\prod_{t'<t}^{t} e^{-w\phi(\mathbf{x},t') \Delta t}= e^{-w
\sum_{t'=0}^{t} \phi(\mathbf{x},t') \Delta t}.
\end{equation}

When the continuous limit is taken on the discrete lattice, the \textit{probability of having been infected at least once in
the past} becomes $1-\exp(-w
\int_{0}^{t}\phi(\mathbf{x},t^{\prime})dt^{\prime})$. This expression
is the probability for a site to be immune at
time $t$. Since $\lambda$ is the parameter related to the immunization in the system, we can use the
above result to write an expression for $p(\mathbf{x},t)$,
\begin{equation}
\label{p(r,t)}
p(\mathbf{x},t) = p_0 - \lambda \left[ 1-e^{\textstyle - w\int_{0}^{t} \phi(\mathbf{x},t^{\prime}) dt^{\prime}} \right].
\end{equation}
We assume that the mass parameter $r$ can be written as $r \propto p_c -
p(\mathbf{x},t)$. Thus, using Eq.~\ref{p(r,t)}, the addition of
immunization can be reflected in the action as a modification in the
original DP mass parameter by the following substitution:

\begin{equation}
r \rightarrow r + \lambda \left[1-e^{\textstyle - w \int_{0}^{t}
\phi(\mathbf{x},t^{\prime})dt^{\prime}} \right].
\end{equation}

Finally the action for DP with
immunization can be written as a modification of the directed
percolation action as follows:
\begin{eqnarray}
\label{NonMarkovianAction}
S &=& S_{\rm DP} + \nonumber \\
&\lambda& \int dt d^{d}x
\tilde{\phi}(\mathbf{x},t)\phi(\mathbf{x},t) \left[1- e^{ - w
\int_{0}^{t}dt^\prime \phi(\mathbf{x},t')}\right].
\end{eqnarray}

According to the field theory, $w$ is a finite coupling constant. Then in
the numerical calculations on a discrete $d+1$ dimensional lattice, $w$ can be considered as a parameter such
that the subsequent infection rate is,
\begin{equation}
p_0+(p-p_0) \left[1-\exp(-w n(\mathbf{x},t))\right].
\end{equation}
Here, the function $n$ counts all the past activity at a site $\mathbf{x}$
until time $t$. Since the exponential function vanishes at sufficient
large times, the \textit{effective} subsequent infection rate is
equal to $p$.

In the numerical calculations carried out in the present work,  we assume that the susceptibility
for spreading only changes at the first infection process and remains
constant thereafter. This assumption is equivalent to take an
infinite value of $w$, but does not make any relevant change in the
final results of the theory as it is argued in Ref.~\cite{GCR97}.
%
%
%
\subsection{Renormalization group analysis}
\label{Renormalization}

\noindent

\begin{figure*}
\includegraphics[width=150mm]{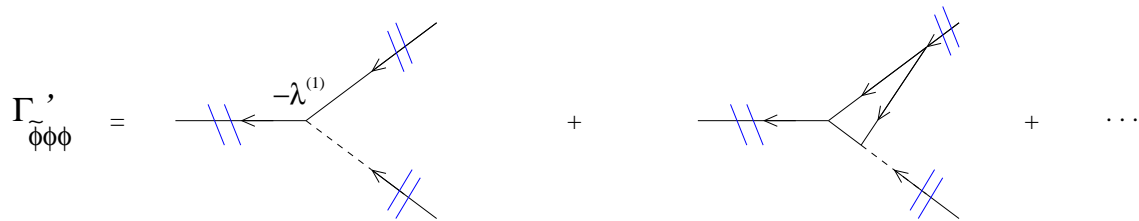}
\caption{
\label{FIGgammaprime}
One-loop diagrams for the three point vertex function
$\Gamma_{\tilde{\phi}\phi\phi}^{\prime}$. The dashed lines in the external
dashed-continue-leg represent the shift in time indicated by the
$\Theta(t-t^{\prime})$ function in the action after the expansion of
the exponential. The 'external legs' are cut off,
and this fact is indicated with the two parallel lines. 
} 
\end{figure*}      
\begin{figure*}
\includegraphics[width=150mm]{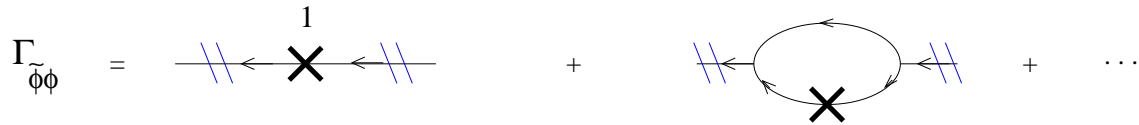}
\caption{
\label{FIGgamma2}
One-loop diagrams for the two point vertex function
$\Gamma_{\tilde{\phi}\phi}$ . The 'external legs' are cut off,
and this fact is indicated with the two parallel lines. The
\textsf{\textbf{X}} indicates the insertion of the composite operator $\tilde{\phi}\phi$. 
} 
\end{figure*}

Expanding in Eq.~\ref{NonMarkovianAction} the exponential function as a power series, the action reads
\begin{eqnarray}
\label{ExpandedAction}
S & = & S_{\rm DP} + \nonumber \int dt d^{d}x
\tilde{\phi}(\mathbf{x},t)\phi(\mathbf{x},t)  \times \nonumber \\
& & \biggl[\,-\sum_{n=1}^\infty \frac{\lambda^{(n)}}{n!} \left(-\int_{0}^{t}dt' \phi(\mathbf{x},t') \right)^n \biggr],
\end{eqnarray}
where we abbreviate $\lambda^{(n)}= \lambda w^n$. Let us now study the stability of
the DP fixed point with respect to a small perturbation in
$\lambda^{(n)}$ by dimensional analysis. Introducing a length scale
$k^{-1}$, the mean-field fixed point of DP is characterized by the
dimensions $[\tilde{\phi} \phi]=k^{d}$, $[D]= \omega k^{-2}$, and
$[u/D]=k^{4-d/2}$ so that the upper critical dimension is
$d_c=4$. When immunization is considered as a perturbation,  the
effective expansion parameter in the expression of the two point
vertex function is $u_2^n \lambda^{(n)}$. Thus, by dimensional analysis we find
\begin{equation}
[\frac{u_2^n \lambda^{(n)}}{D_0^{2n+1}}]=k^{2+n(4-d)} \,.
\end{equation}
Consequently the upper critical dimension for each coupling constant $\lambda^{(n)}$ is
$d_c=\frac{2}{n}+4$, where $n=1,\ldots,\infty$. This means that for dimensions $d>6$,  $\lambda^{(n)}$ is
irrelevant for all $n$. The parameter $u_2$ will also be irrelevant, and 
only the mass parameter $r$ renormalizes.
 
Consider now the renormalization group (RG)
parameter space spanned by $r$, $u_2$, and the $\lambda^{(n)}$. The DP fixed point occurs at a point on the $u_2$ axis, and
the GEP fixed point is somewhere in the hyper-dimensional
space. Since all the coupling constants are irrelevant in the case
$d>6$, the RG flows bring the system from the neighborhood of the DP fixed
point towards the Gaussian DP fixed point, which corresponds to the DP free-field theory.
Therefore if close to the DP fixed point we turn on a small perturbation of
immunization in a system, the qualitative
nature of the system will not change, and the epidemic process is
expected to spread according to a mean-field DP process controlled
by the probability of first infections. 

For $4 < d \leq 6$, the coupling constant $\lambda^{(1)}$ is the most
relevant. Consequently the higher order terms in the expansion of the exponential
in Eq.~(\ref{ExpandedAction}) can be neglected. In this case the RG
flows bring the system away from the neighbourhood of the DP fixed point
to the GEP fixed point which is stable. Thus, we expect the system to
undergo an ordinary crossover from DP to the dynamic percolation
universality class at a certain typical time scale. However, recently
it has been suggested that dangerous irrelevant operators may possibly
lead to a non-trivial critical behavior different from dynamical
percolation~\cite{FvW02}.

To take the fluctuation effects into account, we define the
renormalized $\lambda^{(1)}$ as $\lambda_R^{(1)}$ and we apply  standard
methods of the perturbative renormalization group
~\cite{Amit78}. In particular we make use of an $\epsilon=4-d$ expansion around the DP
critical point. The non-Markovian modification in the DP action can be
written then as an additional term of the form,

\begin{equation}
\label{lambda1term}
\lambda^{(1)} \int dt \, d^{d}x \,\, \Theta(t-t') \tilde{\phi}(\mathbf{x},t) \phi(\mathbf{x},t) \phi(\mathbf{x},t').
\end{equation}
The non-locality in time, expressed by the $\Theta$ function
in Eq.~\ref{lambda1term}, is represented by a dashed line propagator in the
Feynman diagrams. The renormalized coupling constant $\lambda_{R}^{(1)}$ can
be determined in terms of the vertex function
$\Gamma_{\tilde{\phi}\phi\phi}^{\prime}$ (see Fig.~\ref{FIGgammaprime}), evaluated at some specified
normalization point, NP, setting a momentum scale $\mu$. The $ ^\prime$
in the vertex function's notation in Eq.~\ref{lambda1term} is used to
indicate that this function is calculated in the DP with immunization theory and is
different from the one calculated in the ordinary DP. Near the upper
critical dimension the ultraviolet divergences are absorbed in the
renormalization constant $Z_{\phi}$, where $\phi_{R}\equiv
Z_{\phi}^{-1/2} \phi$ and $Z_{\phi}=Z_{\tilde{\phi}}$.
\begin{eqnarray}
\label{EqlambdaR}
\lambda_{R}^{(1)} & = & - \Gamma_{R \tilde{\phi}\phi\phi}^{\prime} \biggr|_{NP}\nonumber \\
            & = & - Z_{\phi}^{3/2}\Gamma_{\tilde{\phi}\phi\phi}^{\prime}\biggr|_{NP}.
\end{eqnarray}
The structure of the corrections to the vertex function
$\Gamma_{\tilde{\phi}\phi\phi}^{\prime}$ suggests a correspondence with the
RG theory of DP away from the critical point, where is considered
the renormalization of a nonzero mass term $r$ which couples with the composite operator
$\tilde{\phi}(x,t)\phi(x,t)$. One loop corrections to the
corresponding two point bare vertex function
$\Gamma_{\tilde{\phi}\phi}$ in this theory, are depicted in
 Fig.~\ref{FIGgamma2}. The normalization condition implies that the renormalized two
point vertex function $\Gamma_{R \tilde{\phi}\phi}$ evaluated at the
normalization point NP, is equal to 1.
\begin{equation}
\label{EqgammaR}
\Gamma_{R \tilde{\phi}\phi}\biggr|_{NP}=Z_{\tilde{\phi}\phi}^{-1} Z_{\phi}\Gamma_{\tilde{\phi}\phi}\biggr|_{NP}=1.
\end{equation}
Comparing these corrections with those appearing in the expression to
$\Gamma_{\tilde{\phi}\phi\phi}^{\prime}$, it is possible to see that the insertion of the composite operator 
$\tilde{\phi}\phi$ is equivalent to the ``insertion'' of the
dashed-continuous-line leg in Fig.~\ref{FIGgammaprime}. By inspection of all possible Feynman diagrams,
we conclude that this is valid to all orders in perturbation
theory. Therefore the Feynman integrals involved  in both $\lambda^{(1)}$
and $r$ renormalizations are identical, and we can write:
\begin{equation}
\label{Eqanalogy}
 - \lambda^{(1)} \Gamma_{\tilde{\phi}\phi} =  \Gamma_{\tilde{\phi}\phi\phi}^{\prime}.
\end{equation}
Consequently from Eq.~(\ref{EqlambdaR}),~(\ref{EqgammaR}) and ~(\ref{Eqanalogy}), it can be proved that,
\begin{equation}
\label{lambdaEND}
\lambda_{R}^{(1)} = \lambda^{(1)}  Z_{\phi}^{1/2} Z_{\tilde{\phi}\phi}.
\end{equation}
To describe how $\lambda_{R}^{(1)}$ flows under renormalization, it is
necessary to define a Callan-Symanzik beta function. The
dimensionless coupling constant corresponding to $\lambda_{R}^{(1)}$ is
\begin{eqnarray}
g_{R}^{\prime} & = & \biggl(\frac{\lambda_{R}^{(1)}}{D_{R}^{2}}\biggr)
\mu^{-2-\epsilon/2} \nonumber \\
& = & \biggl(\frac{\lambda^{(1)}}{D_{0}^{2}}\biggr)\mu^{-2-\epsilon/2} Z_{\phi}^{1/2} Z_{\tilde{\phi}\phi}Z_{D}^{-2},
\end{eqnarray}
where $\epsilon=4-d$, and $D_{R}=D_{0}Z_{D}$ is the renormalized
diffusion coefficient. Then the beta function, which gives the
differential renormalization group flow equation of $g_{R}$ is
\begin{equation}
\beta(g_{R}^{\prime}) = \mu \frac{\partial g_{R}^{\prime}}{\partial \mu} =  g_{R}^{\prime} [-2-\frac{\epsilon}{2}+\frac{1}{2}\gamma_{\phi}+\gamma_{\tilde{\phi}\phi}-2\gamma_{D}],
\end{equation} 
where $\gamma_{\phi}$, $\gamma_{\tilde{\phi}\phi}$ and $\gamma_{D}$
are defined as
\begin{equation}
\gamma_{\phi}=\mu \frac{\partial ln Z_{\phi}}{\partial\mu},
\end{equation}
\begin{equation}
\gamma_{\tilde{\phi}\phi}=\mu \frac{\partial ln Z_{\tilde{\phi}\phi}}{\partial\mu}, 
\end{equation}
\begin{equation}
\gamma_{D}=\mu \frac{\partial ln Z_{D}}{\partial\mu}.
\end{equation}
At the DP fixed point these gamma functions are related to the
critical exponents by
$\gamma_{\phi}^{*}=\frac{2\beta}{\nu_{\perp}}-d$,
$\gamma_{\tilde{\phi}\phi}^{*}=z-\frac{1}{\nu_{\perp}}$ and
$\gamma_{D}^{*}=z-2$, and thus 
\begin{equation}
\beta(g_{R}^{\prime}) = \frac{1}{\nu_{\perp}}(\beta-\nu_{\parallel}-1)g_{R}^{\prime}+O(g_{R}^{\prime
2}).
\end{equation}
Then, as we are approaching the infrared limit
$(k \rightarrow 0)$, $g_{R}^{\prime}$ increases, being relevant to all orders in
perturbation theory.

The renormalization group eigenvalue $y_{\lambda^{(1)}}$, corresponding to
$\lambda^{(1)}$, is equal to
\begin{equation}
\label{YLambda1}
y_{\lambda^{(1)}} = \frac{1}{\nu_{\perp}}(1+\nu_\parallel-\beta) \,.
\end{equation}
Since this expression is always positive, the first term of the
expansion in Eq.~(\ref{ExpandedAction}) is relevant to all orders of
perturbation theory around $d=d_c=4$.

However, for $d<d_c=4$ the coupling constants are increasingly relevant in the expansion~(\ref{ExpandedAction}), and it is no longer meaningful to expand the exponential
function. The system is driven away from the vicinity of the DP
fixed point towards a non-trivial fixed point of order $4-d$. To study
this case, we apply a scaling analysis. Assume that position
and time scale as $[x]=k^{-1}$
and $[t]=k^{-z}$. In the DP theory
away from criticality, the non-zero mass term scales as
$[r]=k^{1/\nu_{\perp}}$. It couples to the operator
$\tilde{\phi}\phi$, which has a scaling dimension
$x_{\tilde{\phi}\phi}$, such that,
$[\tilde{\phi}\phi]=k^{x_{\tilde{\phi}\phi}}$. Using the fact that the
action has to be dimensionless, we find that 
$x_{\tilde{\phi}\phi}=d+z-\frac{1}{\nu_{\perp}}$. 
Let us focus now on the scaling analysis of the non-Markovian term added to the
DP action. First we consider the term
corresponding to $n=1$. $\lambda^{(1)}$ scales as
$[\lambda^{(1)}]=k^{y_{\lambda^{(1)}}}$. Notice that the operators
$\tilde{\phi}(x,t)\phi(x,t)$ and $\phi(x,t^{\prime})$ are evaluated at
different space-time points, and consequently they are not correlated. So,
each of them scales with its own scaling dimension, $x_{\tilde{\phi}\phi}$ and $x_{\phi}=\beta/\nu_{\perp}$, such that,
$[\tilde{\phi}(x,t)\phi(x,t)]=k^{x_{\tilde{\phi}\phi}}$,
$[\phi(x,t^{\prime})]=k^{x_{\phi}}$. From the dimensionless nature
of the action, we thus obtain $y_{\lambda}=-\frac{1}{\nu_{\perp}}(\beta-\nu_{\parallel}-1)$.
Generalizing this for  higher-order terms in the
exponential expansion we find that the scaling exponent for each $\lambda^{(n)}$ are,
\begin{equation}
\label{ylambdan}
y_{\lambda^{(n)}} = 
\frac{1}{\nu_{\perp}}\bigl(1+n(\nu_\parallel-\beta)\bigr)\,.
\end{equation}
Since $\nu_\parallel-\beta$ is always positive in $d<4$ dimensions,
the terms of the expansion~(\ref{ExpandedAction}) are still
increasingly relevant.

Furthermore, the scaling invariance of the non-Markovian term can only
be established if the exponential function and its arguments are
dimensionless. Therefore the two couplings $w$ and
$\lambda$ scale separately with different scaling
exponents, i.e., $[\lambda]=k^{y_{\lambda}}$ and  $[w]=k^{y_{w}}$. From
Eq.~(\ref{ylambdan}) we obtain,
\begin{equation}
\label{ylambdayw}
y_\lambda = \frac{1}{\nu_\perp}\,, \qquad y_w = \frac{\nu_\parallel-\beta}{\nu_\perp}\,.
\end{equation}
Consequently the coupling constant $w$, is also relevant under renormalization. 
\begin{figure*}
\includegraphics[width=130mm]{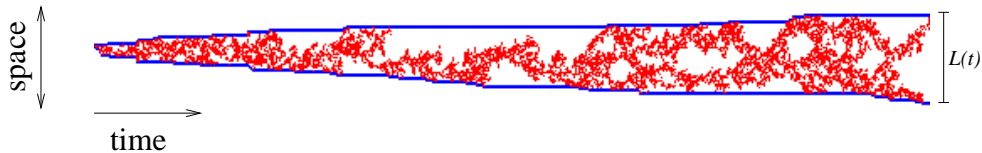}
\caption{
\label{FIGLOW}
Spreading in the limit of a very small first infection probability
$p_0$. The figure shows a surviving cluster in 1+1 dimensions. In this
case the domain of infected/immune is hole-less and grows very slowly
with time. Its boundaries can be considered as stationary absorbing walls.
} 
\end{figure*}      
%

\section{Low first infection probability limit}
\label{LOW}

In the previous section using scaling arguments, we showed that the non-Markovian term is
relevant for $d<d_c=4$. We can now argue that a critical
spreading process on the horizontal line in Fig.~\ref{FIGPHASEDIAG}, is
therefore driven away from the vicinities of the DP critical point to the points $A$
or $B$, where the probability of first infections is either very low
or very high.
In order to understand the dynamic
behavior in these limits we study the asymptotic spreading properties
keeping the second infection probabilities  at the critical value of
DP. We consider an initial state with a single infected site at the
origin in a non-infected environment. For simplicity we will focus on
the $1+1$-dimensional lattice. In this case there are not empty
sites inside the epidemic clusters, which are
composed of infected and immune sites only. 
%
\subsection{Quasi-static approximation}
\label{quasi-stat}
%
Let us consider the limit of a very small first infection probability. In this
regime, although first infections hardly ever
happen, there still exists a surviving critical epidemic process as
is shown in Fig.~\ref{FIGLOW}. The active domain is composed by re-infected sites
among immune ones, and can be considered as bounded by almost
rigid ``walls'' of empty sites. Let us define $L(t)$ to be the
distance at time t between
these two walls, such that there is at least one site active. If we
keep the second infection rate critical, the 
 dynamic epidemic process can be thought as an effective critical DP process evolving in a finite system of
size $L(t)$. 

A quasi-static approximation is based on the assumption that the time scale on
which $L(t)$ grows is much larger than the correlation time of the DP
process. Therefore,  $L(t)$ can be assumed constant.
It is known that in a finite-size system with a constant width $L$, the
survival probability of a DP process decays exponentially as $P(t)
\sim \exp(-t/L^{z})$, where $z=\frac{\nu_\parallel}{\nu_\perp}$. Then, in
a quasi-static finite system the survival probability can be
approximated as $P_s(t) \sim \exp(-t/\xi_\parallel)$, i.e.
\begin{equation}
\label{PEquation}
\frac{d P_s(t)}{dt} \;\simeq\; -\frac{1}{\xi_\parallel}P_s(t) \;\simeq\; -a \,P_s(t)\,L(t)^{-z}\,,
\end{equation}
where $a$ is a non-universal amplitude factor. We need now to know how $L(t)$ grows. Since the two boundaries are absorbing, we can apply the theory of DP with absorbing walls~\cite{FHL01}. According to these results, the density of active sites next to the walls generated by the surviving clusters scales as
\begin{equation}
\rho_s(t) \;=\; b\,L(t)^{-\beta_s/\nu_\perp} \,,
\end{equation}
where $b$ is a non-universal amplitude factor which depends on the
value of $p_0$. $\beta_s \simeq 0.73371(2)$ is the {\em surface critical exponent} of DP in 1+1 dimensions.
Clearly, $dL/dt$ is proportional to the frequency by which the system attempts to infect sites at the boundary and thus also to the density of active sites next to the boundary:
\begin{equation}
\label{LEquation}
\frac{ dL(t)}{dt} \;=\; p_0\; \rho_s(t)  \;=\; p_0\, b\, L(t)^{-\beta_s/\nu_\perp} \,.
\end{equation}
Therefore, the size of the domain grows as
\begin{equation}
\label{DomainSizeGrowth}
L(t) \;=\; (p_0 \,b\,(1+\beta_s/\nu_\perp) \, t)^{1/(1+\beta_s/\nu_\perp)} .
\end{equation}
Inserting this result into Eq.~(\ref{PEquation}) and solving the differential equation we obtain
\begin{equation}
\ln P_s(t) \;\sim\; - \int_0^t L(z)^{-z} dt \;\sim\; -p_0^{-\alpha}\, t^{1-\alpha}\,,
\end{equation}
where $\alpha = \frac{\nu_\parallel}{\nu_\perp+\beta_s}\simeq
0.947167$. Hence the survival decays asymptotically as a stretched exponential of the form
\begin{equation}
\label{StretchedExponential}
P_s(t) = P_s(0) \exp \left( -A \,p_0^{-\alpha} \, t^{1-\alpha} \right)\,.
\end{equation}
with $A = \frac{a}{1-\alpha} (\frac{\alpha}{bz})^\alpha$. The above result implies that the average survival time $T$ is {\em finite} and scales as
\begin{equation}
T \sim p_0^{\alpha/(1-\alpha)}.
\end{equation}
It is interesting to compare these results with recent findings
for random walkers between movable reflectors, where the survival
probability was shown to decay as a power law with continuously 
varying exponents~\cite{Dickman}. We note that this result is not 
in contradiction with the present work since it would correspond 
to the limit $\alpha\to 1$.
%
\subsection{Numerical simulations}
%
The lattice model considered is a 1+1 dimensional triangular lattice. We
simulate a directed bond percolation process with a first infection rate $p_0$ different from  a rate for
subsequent infections or second infection rate $p$.

\begin{figure}
\includegraphics[width=85mm]{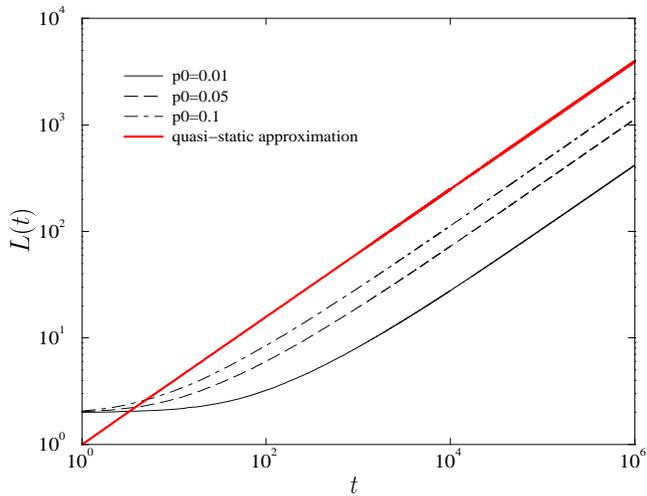}
\caption{
\label{FIGL}
The logarithm of the domain size $L(t)$ in the problem of directed percolation with immunization for
small values of $p_0$ as function of time $t$. The numerical results
are compared with the quasic-static approximation predictions.
} 
\end{figure}

\begin{figure}
\includegraphics[width=85mm]{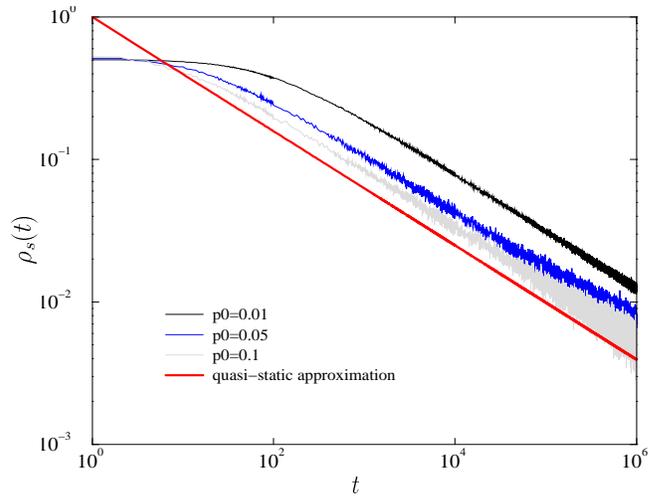}
\caption{
\label{FIGdens}
The logarithm of the surface
density of active sites of directed percolation with immunization for
small values of $p_0$ as function of time $t$.  The numerical results
are compared with the quasic-static approximation predictions.
} 
\end{figure}

Let us first consider the limit of a very small first infection
probability, where it is difficult for the process to infect
sites at the boundary that have never been infected before (see Fig.~\ref{FIGLOW}). Bonds inside the infected region
are open with the critical probability
$p=p_c=0.6447001(1)$~\cite{Jensen96}, while bonds leading to healthy
sites at the boundary are open with a different probability of first
infection $p_0$.  However, as $p_0$ is very small, conventional seed
simulations are not suitable since most of the runs terminate after
a very short time. For example, for $p=0.1$ most of the runs survive for
less than $100$ time steps. Consequently we apply enrichment methods in order to circumvent this limitation. 
First of all, we apply a very simple enrichment method, in which we
consider one lattice system. We keep the activity artificially
alive at any time $t$. This is achieved by ignoring the updates which
lead to the inactive state. Averaging over
many realizations we measured the  domain size $L(t)$ and the surface
density of active sites. We find that the exponents predicted by the quasi-static
approximation are in good agreement with the values obtained with the
numerical simulations as shown in Fig.~\ref{FIGL} and
Fig.~\ref{FIGdens}. The nonuniversal factor $b$ in
Eq.~(\ref{DomainSizeGrowth}) depends on the value of $p_0$. For,
$p_0=0.01$ it takes the value $b=1.65$.

\begin{figure}
\includegraphics[width=85mm]{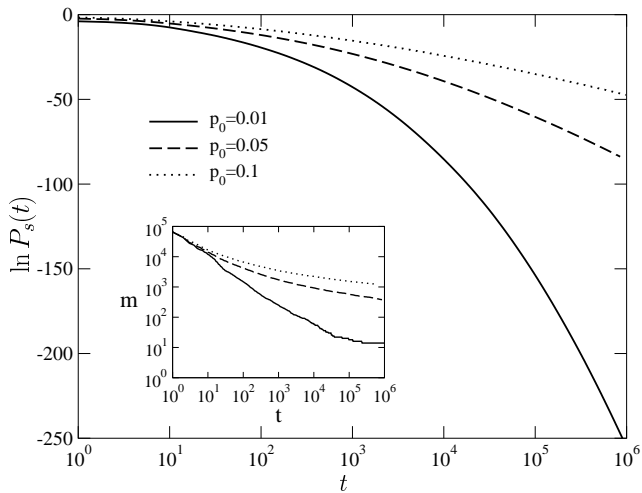}
\caption{
\label{FIGSURV}
The logarithm of the survival probability $P_s(t)$ of directed percolation with immunization for small values of $p_0$. The inset shows the corresponding number of independent ancestors as a function of time (see text). For $p_0=0.01$ the duplication method reaches its limit since only $11$ independent ancestors are left.
} 
\end{figure}

To determine the survival probability $P_s(t)$, we apply another
enrichment method~\cite{GrassWinners}, which leads to a considerable
improvement. The simulation starts with an {\it
ensemble} of $N=65\,536$ independent systems. Whenever the number of
active runs becomes smaller than $N/2$ the ensemble is duplicated by
creating identical copies of all the remaining active states. The new
systems are labelled according to their ancestors. The survival
probability $P_{s}$ then is reduced by a factor of $n_f/n_i$, where $n_{i}$ is the number of
initial active systems and $n_{f}$ is the number of remaining active
systems before the duplication. This process may be repeated as long as the ensemble has a
sufficiently large number $m$ of independent ancestors at
$t=0$. Using this method we were able to extend the temporal range of
the simulation by four orders of magnitude up to $t=10^6$. Our results
are shown in Fig.~\ref{FIGSURV} for various values of $p_0$. As can be
seen, the survival probability plotted in a double logarithmic
representation is not a straight line, proving that it does {\em not}
follow a power law. For the case $p_0=0.01$, the parameter $a$ in
Eq.~(\ref{PEquation}) takes the value $a = 1.00(2)$. 
%
%
\subsection{Lattice effects}
%
The scaling arguments in Section~\ref{quasi-stat} are
developed assuming the existence of a well defined mean value for
the domain size $L(t)$. To justify this assumption, we define $P(L,t)$
to be the probability of having a domain of size $L$ at time $t$. Then
the survival probability is $P_s(t)=\int P(L,t)dL$. In this subsection
we show that $L(t)$ is peaked around the mean value $L(t)$.

We begin by writing down a heuristic discrete master equation for the temporal evolution
of $P(L,t)$:
\begin{eqnarray}
\label{MEP}
P(L,t+1) & = & e^{-a L^{-z}} P(L,t) + b p_0
(L-1)^{-\beta_s/\nu_\perp} \times \nonumber \\
& & P(L-1,t) - b p_0 L^{-\beta_s/\nu_\perp} P(L,t).
\end{eqnarray}
The first term in the right hand side of Eq.~(\ref{MEP}) describes the change in $P(L,t)$ due to terminating
runs according to Eq.~(\ref{PEquation}) integrated over the time
interval $[t,t+1]$. The second term corresponds to the probability of
creating a domain of size $L$ from a domain of length ($L-1$). The third
term describes a loss term corresponding to $L\rightarrow L+1$.
We neglect the second order contributions $+ b^{2} p_0^{2} 
(L-1)^{-2\beta_s/\nu_\perp} P(L-2,t)$ and $-b^{2} p_0^{2} 
(L-1)^{-2\beta_s/\nu_\perp}$ which only affect the behaviour at small time and
small $L$ regime.

It is straightforward to show that this master equation is consistent with
Eq.~(\ref{PEquation}), by summing over all $L$ and replacing $L$ by
its average value $L(t)$. On doing this we
obtain $d P_s(t)/dt = -a L(t)^{-z} P_s(t)$, that is the same
equation as Eq.~(\ref{PEquation}).

Consider Eq.~(\ref{MEP}) with $p_0$, $a$ and $b$ fixed at certain values. Then the differential equation
which corresponds to the master equation, Eq.~(\ref{MEP}), can be
written as
\begin{eqnarray}
\label{MEPa}
\frac{\partial P(L,t)}{\partial t} & = & - \frac{a}{L^z} P(L,t) -
\nonumber \\
& & p_0 b\frac{\partial}{\partial L}\left[ L^{-\beta_s/\nu_\perp}
P(L,t)\right], \, L \gg 1.
\end{eqnarray}
The solution for Eq.~(\ref{MEPa}) is
\begin{eqnarray}
\label{Pasol}
P(L,t) & = & e^{-\frac{a}{p_0 b(1-z+\beta_s/\nu_\perp)}
L^{1-z+\beta_s/\nu_\perp}} L^{\beta_s/\nu_\perp} \nonumber \\ & & 
\psi \left[ L^{\beta_s/\nu_\perp+1} - p_0 b (\beta_s/\nu_\perp+1) t\right],
\end{eqnarray}
where $L \gg 1$. Consider the integrated distribution probability $G(L,t)=\int_L^\infty
P(L,t) dL$. Then $G(L,t)$ has the functional form
\begin{eqnarray}
\label{Ga}
G(L,t) &=& e^{-c \left[ p_0 b (\beta_s/\nu_\perp +1) t
\right]^\kappa} \times \nonumber \\
 & & \phi\left[ L^{\frac{\beta_s}{\nu_\perp}+1} - p_0 b (\frac{\beta_s}{\nu_\perp}+1)
t \right] \left(1+ O(\frac{1}{t^{1-\kappa}})\right), \nonumber \\
 & &       
\end{eqnarray}
with $c=a/(p_0 b(1-z+\beta_s/\nu_\perp))$ and
$\kappa=(1-z+\beta_s/\nu_\perp)/(1+\beta_s/\nu_\perp)$.

\begin{figure}
\includegraphics[width=87mm]{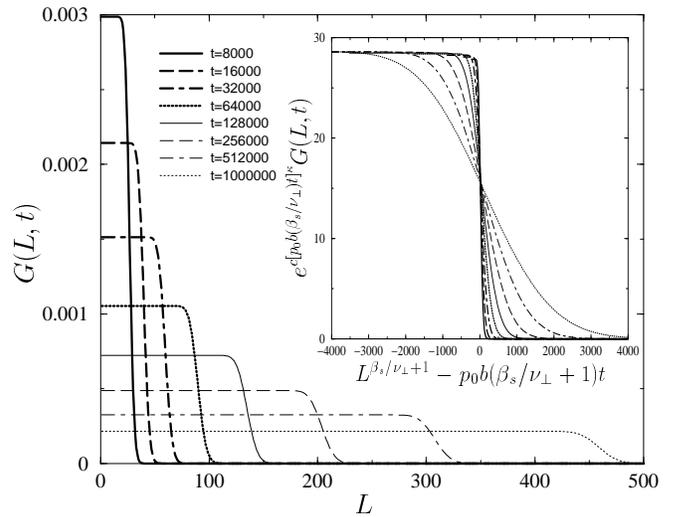}
\caption{
\label{FIGGa001}
Probability $G(L,t)=\int_L^\infty P(L,t) dL$ computed by numerical integration
of the master equation Eq.~(\ref{MEP}). The plots correspond to
$a=0.01$ and $p_0=0.01$. The inset shows the scaling plot when $G(L,t)$ scales according to Eq.~(\ref{Ga}).  
}
\end{figure}

To observe the qualitative behaviour of $P(L,t)$ and $G(L,t)$, we
carry out numerical integrations of Eq.~(\ref{MEP}). As initial conditions we
choose the state in which there is only one site infected at time $t=0$.
In Fig.~\ref{FIGGa001}, $G(L,t)$ is shown for several values of
time $t$. It behaves like a propagating front with an overall
exponential factor of time $t$. The fact that all the $G(L,t)$ curves  intersect at the
same point (see inset in Fig.~\ref{FIGGa001}), implies that there is a
well defined peak and a mean value $L(t)$ for $P(L,t)$. The
qualitative behaviour of $G(L,t)$ is independent of the particular
values of $a$ and $p_0$. Consequently, to perform these calculations,
we use values of $a$ and $p_0$ such that the effect of the exponential
decay in Eq.~(\ref{Ga}) is not so pronounced.
\begin{figure}
\includegraphics[width=85mm]{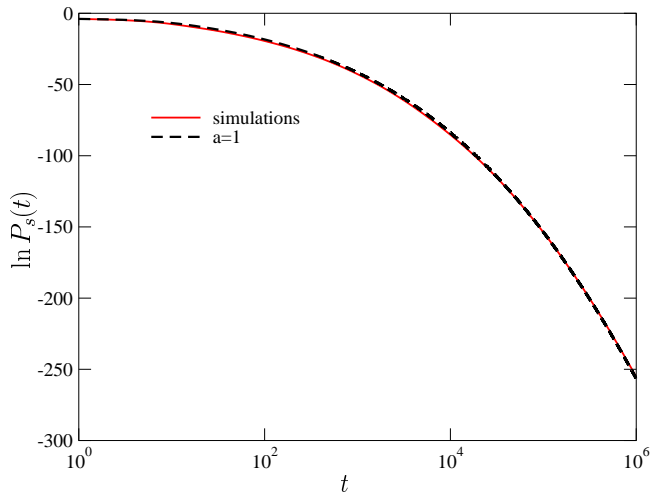}
\caption{
\label{FIGMEsim}
Survival probability calculated using the master equation
Eq.~(\ref{MEP}). We set $p_0=0.01$, $a=1$ and $b=1.65$. The result is in good agreement
with the Monte Carlo simulations for $p_0=0.01$. 
}
\end{figure}

We now compare the predictions of the master equation
proposed in this subsection, with the results obtained from a
simulation of the model. In Eq.~(\ref{MEP}), we fixed the value of
$p_0=0.01$ and vary $a$. The survival probability calculated from direct numerical integration of
Eq.~(\ref{MEP}), describes well the corresponding  simulation result
for $p=0.01$, in the case $a=1$  (see Fig.~\ref{FIGMEsim}).
%
\section{High first infection probability limit}
\label{HIGH}
\begin{figure}
\includegraphics[width=88mm]{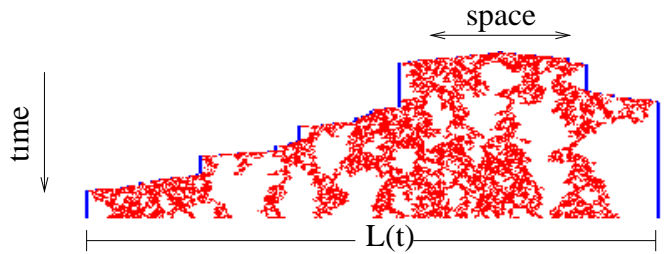}
\caption{
\label{FIGHIGH}
Spreading in the limit of a very high first infection rate. 
} 
\end{figure}

%
Let us finally consider the limit of a very high first infection
probability $p_0 \to 1$. Again we restrict to the 1+1-dimensional
case, where the region of immune sites does not contain healthy sites
inside. Because of the enhanced spreading probability at the boundaries, the domain grows rapidly by sudden {\it avalanches} of successive first infections (see Fig.~\ref{FIGHIGH}). Since the avalanches at the left and the right boundary are expected to be uncorrelated, it suffices to study the propagation of one of the boundaries. As in the previous section, we propose simple scaling arguments in order to describe the growth of the infected/immune domain and the survival probability. 
%
\subsection{Independent avalanche approximation}
%
In 1+1-dimensional directed bond percolation with $p=p_c$ and $p_0 \to 1$ an avalanche is caused by a sequence of open bonds at the boundary. Therefore, the avalanche size $\xi$ is distributed exponentially as
\begin{equation}
P(\xi) \;=\; (1-p_0) \,p_0^\xi \;\sim\; e^{-\xi/\bar{\xi}}\,,
\end{equation}
The quantity
\begin{equation}
\bar{\xi}\;=\;-\frac{1}{\ln p_0} \;\approx\; \frac{1}{1-p_0} \qquad \qquad p_0 \rightarrow 1.
\end{equation}
is the average distance by which the avalanche advances the boundary
in space. After each avalanche the process continues to evolve as an ordinary critical DP process inside the immune domain until it terminates or returns to the boundary where it releases a new avalanche. Thus the spreading behavior is mainly determined by the distribution of waiting times $\tau$ between the avalanches.
We argue that the distribution of waiting times between avalanches is related to the problem of {\it local persistence} in
DP~\cite{HinrichsenKoduvely,DuesseldorfPersistence} (for a recent
review on persistence see e.g.~\cite{Satya99}). The local persistence
probability $R(t)$ is defined as the probability that a randomly
selected site in an ordinary critical DP process starting from a
homogeneous initial state has {\em not} been reactivated until time
$t$. It was shown that this quantity decays algebraically as $R(t)
\sim t^{-\Theta}$, where $\Theta=1.50(1)$ is the so-called local
persistence exponent~\cite{HinrichsenKoduvely}. In the present problem
the situation is similar: Each avalanche creates locally a
quasi-homogeneous state. The process then evolves as an ordinary
critical DP process inside 
the infected/immune region until the boundary is revisited for the first time in order to release a new avalanche. However, unlike persistence studies in 1+1 dimensions, where a persistent site can be activated independently from the left and from the right, the boundary sites in the present problem can be infected only from one side. Hence the probability that the next avalanche has not yet been released decays as $\tau^{-\Theta/2}$. Thus we conjecture that the waiting times between avalanches are distributed algebraically as
\begin{equation}
P(\tau) \;\sim\; \tau^{-1-\Theta/2}\,.
\end{equation}
Next, we argue that these waiting times may be interpreted as directed L\'evy flights in time~\cite{LevyGeneral}. After each flight the domain size $L(t)$ grows on average by the mean avalanche size $\bar{\xi}$. This type of growth may be described by the equation
\begin{equation}
D_t^{\Theta/2} L(t) \sim \bar{\xi}\,,
\end{equation}
where $D_t^{\Theta/2}$ is a fractional derivative defined through its action in Fourier space $D_t^{\Theta/2}e^{i\omega t} = (i\omega)^{\Theta/2}  e^{i\omega t}$. Simple dimensional analysis leads to the result
\begin{equation}
\label{GrowingDomain}
L(t) \;\sim\; \bar{\xi} \, t^{\Theta/2} \,.
\end{equation}
In order to compute the survival probability, we note that during the waiting time the outermost active site departs from the boundary with an average distance $\ell(\tau) \sim \tau^{1/z}$. Obviously, the process can only terminate if this distance is of the same order as the size of the infected/immune domain $L(t)$. Therefore, we expect the distribution of waiting times between avalanches to be cut off by a maximal waiting time
\begin{equation}
\tau_\text{max} \sim L^z(t) \sim \bar{\xi}^z \,t^{z\Theta/2}.
\end{equation}
Consequently, the probability that the process terminates between two avalanches is given by
\begin{equation}
P_0 = \frac{\int_{\tau_\text{max}}^\infty d\tau\,P(\tau)}
{\int_1^\infty d\tau\,P(\tau)} \sim \tau_\text{max}^{-\Theta/2}
\end{equation}
On the other hand, the cutoff due to terminating runs implies that the average waiting time is finite and scales as
\begin{equation}
\bar{\tau} = \frac{\int_1^{\tau_\text{max}}d\tau\,\tau\,P(\tau)}
{\int_1^{\tau_\text{max}}d\tau\,P(\tau)} = \tau_\text{max}^{1-\Theta/2}
\end{equation}
Therefore, the average loss of the survival probability per unit time is given by $P_0/\bar{\tau} \sim 1/\tau_\text{max}$, i.e.
\begin{equation}
\label{PEquation2}
\frac{d P_s(t)}{dt} \;\sim\;  -P_s(t)/\tau_\text{max} \;\sim\;  -P_s(t)\,\bar{\xi}^{-z}\, t^{-z\Theta/2}\,.
\end{equation}

\begin{figure*}
\includegraphics[width=180mm]{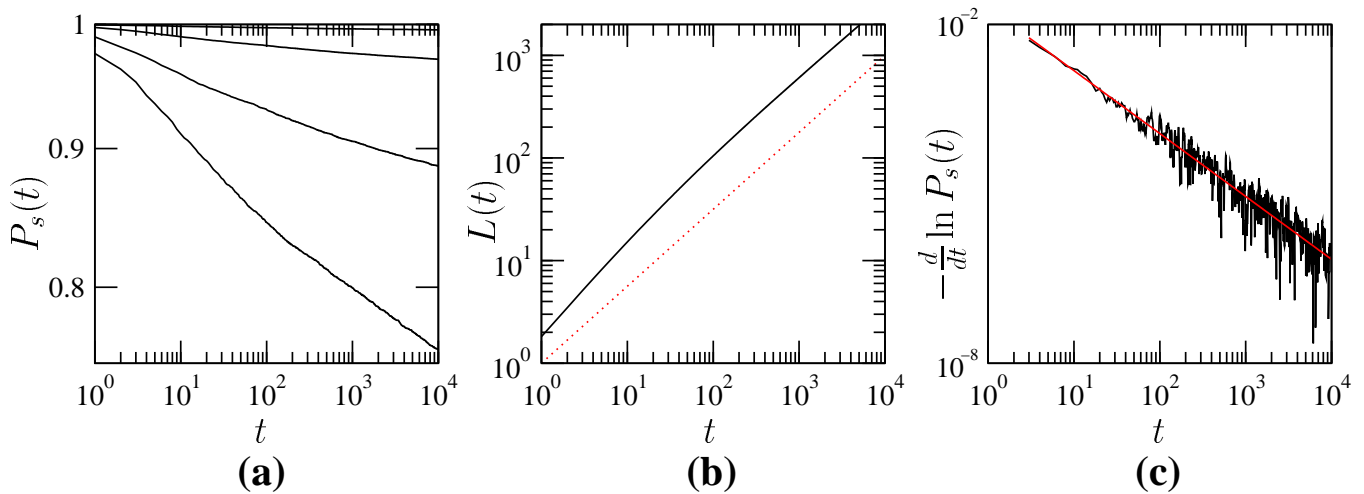}
\caption{
\label{FIGAVALANCHE}
Avalanche approximation in comparison with numerical results. {\bf (a)} Numerically determined survival probability for $p_0=0.85$, $0.9$, $0.95$, and $0.98$ from bottom to top. {\bf (b)} Numerically determined domain size $L(t)$ for $p_0=0.9$ (solid line) compared to the theoretical prediction~(\ref{GrowingDomain}) (dotted line). {\bf (c)} Logarithmic derivative of the survival probability for $p_0=0.9$.} 
\end{figure*}      

Solving this equation, the asymptotic behavior of the survival
probability is not a stretched exponential and is given by a
\begin{equation}
\label{SecondStretchedEx}
P_s(t) = \frac{1}{\tilde{N}} \exp \left(+\tilde{A} \bar{\xi}^{-z}\,t^{1-z \Theta/2} \right)\,,
\end{equation}
where $\tilde{N}$ and $\tilde{A}$ are unknown constants. In contrast to the previous case in Eq.~(\ref{StretchedExponential}) the exponent $1-z \Theta/2  \simeq -0.185$ is negative. Consequently the survival probability tends to a constant $P_s(\infty)>0$, meaning that a finite fraction of runs survives for infinitely long time. For very large $p_0$ this constant is expected to scale as
\begin{equation}
1- P_s(\infty)  \;\propto\; (1-p_0)^{z\Theta/2}\,.
\end{equation}

\subsection{Numerical results}

In order to verify these results numerically, we developed an
especially optimized Monte Carlo algorithm for large values of
$p_0$. In order to compute the survival time for a run
and to see how the boundaries advance for a given realization 
of open and closed bonds, it is in most cases not necessary to construct
the entire cluster. Rather it suffices to construct its branches
next to the boundaries.  For example, in Fig.~\ref{FIGHIGH} large
parts of the cluster are irrelevant for the advancement of the 
boundaries and the survival time. For this reason our algorithm
first constructs the branches of the cluster next to the 
boundaries. If these branches terminate before the 
simulation time is reached, we recursively constructs the omitted branches in the interior
of the clusters. If however they do not terminate, then the recursive
construction is not performed. Using this technique we could extend the simulation time by two decades to $10^4$ time steps, two decades less than in the previous case of low $p_0$. 

Our numerical results are shown in Fig.~\ref{FIGAVALANCHE}. The left
panel shows the survival probability as a function of time in a
directed bond percolation process with $p_0=0.85, 0.90, 0.95,$ and
$0.98$. The positive curvature of the lines indicates that there is no
power-law scaling. Although the upward curvature is in agreement with the expected
 result~(\ref{SecondStretchedEx}), the simulation time is not large enough
 to confirm the behaviour of Eq.~(\ref{SecondStretchedEx}) quantitatively,
 mainly because the constant $\tilde{A}$ is not known. In order to substantiate the assumptions made in the previous subsection, we first measured the growing domain size $L(t)$. As shown in Fig.~\ref{FIGAVALANCHE}b, the measured slope seems to tend to the predicted slope $\Theta/2=0.75(1)$. Moreover, we estimated the logarithmic derivative of the survival probability, in which the unknown prefactor $\tilde{A}$ drops out (see right panel of Fig.~\ref{FIGAVALANCHE}). Although this data set is quite noisy, we observe a rather clean power law. The estimated slope $1.12(10)$ is consistent with the theoretical prediction $t^{-z\Theta/2}=t^{-1.185}$, supporting the assumptions which led to the result~(\ref{SecondStretchedEx}).
%
%
\section{Conclusions}
\label{Conclusions}

We analyzed the effects of immunization as a small perturbation
on the DP model and studied in detail the scaling
behaviour of the theory around the DP critical point. We derived by an
alternative method the field theoretic action
for the model. A non-Markovian term is added to the DP action because
of the presence of immunization.  In the field theory, the probability for
subsequent infections, which is different from the first infection
probability, is a function of position and time. Nevertheless, we assumed in
the lattice model that the susceptibility for spreading changes only
after the first infection, remaining constant thereafter. This
assumption does not changes the final results.

The phase diagram (cf. Fig.~\ref{FIGPHASEDIAG}) comprises two phase
transition lines, namely, a horizontal line, where the process in
reinfected regions shows the critical behavior of DP, and (in more
than one spatial dimension) a curved transition line, where the
critical behavior corresponds to the general epidemic process studied
in~\cite{GCR97}. Both lines meet at the DP critical point $p=p_0=p_c$.

The non-Markovian term turns out to be relevant for $d<6$. We
considered the RG flows and argued that any point in the
neighbourhood of the DP critical point will
be driven away from it. 
In particular, we focused our study on the horizontal
phase transition line with a critical reinfection
probability $p=p_c$. The system along this line, is driven away from
the DP critical point as soon as the immunization effects are turned on
as a small perturbation. The asymptotic behavior is determined by the
limits of very low and high first infection probability (close to
points $A$ and $B$ in Fig.~\ref{FIGPHASEDIAG}).

We proposed simple scaling arguments for the behaviour of the survival
probability in both limits. We related the corresponding exponents with
those exponents of the critical ordinary DP theory and DP near a
wall. The survival probability is found to obey a stretched
exponential behaviour in the low first infection probability limit,
and it decays to a constant in the high first infection probability
limit, taking the form
\begin{equation}
P_s(t) \propto
\begin{cases}
\exp(-at^{+0.0528}) & \text{ for } p_0 \to 0 \\
\exp(+bt^{-0.185}) & \text{ for } p_0 \to 1 
\end{cases}
\end{equation}
The numerical simulations in $1+1$ dimensions support these theoretical predictions, ruling out the
possibility of asymptotic power-law scaling in both limits. The
question as to whether the stretched exponential decay of the survival
probability persists in higher dimensions is still open.

Finally, we comment on a possible connection of these results with
the problem of infinitely many absorbing
states.
A typical example of such systems is the pair contact process $2A \to 3A$, $2A
\to \emptyset$, in which solitary particles are not allowed to
diffuse~\cite{Jensen93}. In
Ref.~\cite{Munoz96,LopezMunoz97,Munoz97,MGD98} this model was studied
with an effective Langevin equation and it was inferred that the
survival probability decays as a power law at the transition point,
with continuously varying exponents. However, the field theoretical action studied in ~\cite{Munoz96,Munoz97,MGD98}
is identical to the one studied in this paper. Hence, we conclude
that, according to the analysis presented in this paper, seed
simulations of the pair contact process, at least in $1+1$ dimensions,
should show an stretched exponential behaviour and not a power low decay at the transition point.

\acknowledgments
We want to thank Peter Grassberger for communication of unpublished
numerical calculations. A.J.D. is greatly indebted to John Cardy for enlightening
discussions and suggestions, and for his most careful reading of the
manuscript. A.J.D. would also like to thank to R. Rajesh for
useful comments.

Part of the present computations have been carried out on the 128-node 
Alpha Linux Cluster Engine  ALiCE at Wuppertal University. We thank N. 
Eicker, Th. Lippert  and B. Orth for their assistance.

A.J.D. was supported by CONICET (Argentina), the British Council-Fundaci\'on
Antorchas (Argentina), and the ORS Award Scheme (UK).
%
%
%

\end{document}